# Book-keeping of Ion and Electron Currents in the Point Defect Model and Role of an Electron Channel at the metal/film Interface in Determining the Forms of $\phi_{m/f}$ and $\phi_{f/s}$

Bosco Emmanuel, CSIR-CECRI, Karaikudi-630006, India

## Abstract

In any consistent electrochemical systems model that addresses more than one interface, the total current at the interfaces should be equal to one another and to the current in the external circuit. The point defect model of Macdonald and co-workers fails to recognize this basic requirement. Hence there is no assurance that currents predicted by the PDM will be the same at the metal/film and the film/solution interfaces. This deficiency is corrected in the present work. Besides, The point defect model lacks a proper book-keeping of ion and electron currents and in particular missed the role of a purely electron channel (e-channel) at the metal/film interface [1]. In this work we describe this book-keeping, include the e-channel and show how this affects the forms for the potentials $\phi_{m/f}$ and $\phi_{f/s}$.

## Introduction

The point defect model lacks a proper book-keeping of ion and electron currents and in particular missed the role of a purely electron channel (e-channel) at the metal/film interface [1]. In this work we describe this book-keeping, include the e-channel and show how this affects the forms for the potentials $\phi_{m/f}$ and $\phi_{f/s}$.

The point defect model considers ion and electron transport in the metal-barrier-oxide-solution sandwich. Needless to say that the metal supports only electron current, the barrier oxide layer both ion and electron currents and the solution only ion current. Further the total current at the metal/film inter face should equal that at the film/solution interface.

## Book-keeping of ion and electron currents at the film/solution interface f/s

The ion and electron current densities at the film/solution interface are given by:

Ion current density at f/s = $-\chi.F.J_M^{f/s} + \chi.F.J_i^{f/s} + 2F.J_O^{f/s}$ (1)

Electron current density at f/s = $(\delta - \chi)F(-J_M^{f/s} + J_i^{f/s} + R_7)$ (2)

If $\delta = \chi$, the e-current in the oxide is zero. On the other hand, if $\delta \neq \chi$, this e-current in the oxide will be non-zero and hence we need to include a purely e-channel at the metal/film interface for this e-current to flow from the oxide layer into the metal.

**Book-keeping of ion and electron currents at the metal/film interface m/f**

The ion and electron current densities at the metal/film interface are given by:

Ion current density at m/f = $-\chi.F.J_M^{m/f} + \chi.F.J_i^{m/f} + 2F.J_O^{m/f}$ (3)

Electron current density at m/f = $i_e^{m/f}$ (4)

Equating the ion plus electron current densities at the m/f and f/s interfaces it follows from equations 1 through 4 that

$$i_e^{m/f} = \chi.F(J_M^{m/f} - J_m^{f/s}) - \chi.F(J_i^{m/f} - J_i^{f/s}) - 2F(J_O^{m/f} - J_O^{f/s}) + (\delta - \chi).F(-J_M^{f/s} + J_i^{f/s} + R_7)$$

... (5)

Equation (5) is the most general expression for the current density in the e-channel at the m/f interface which holds good even for non-steady states.

Now, for the steady state, the flux of each defect is the same at the m/f and f/s interfaces. Therefore equation (5) reduces to

$$i_e^{m/f} = (\delta - \chi).F(-J_M^{f/s} + J_i^{f/s} + R_7) \qquad (6)$$

If we assume that the metal/film interface be modeled as a Schottky contact, the current density $i_e^{m/f}$ is given by the Schottky diode equation:

$$i_e^{m/f} = i_{sat}.(\exp[F.(\phi_{m/f} - \phi_{m/f}^0)/RT] - 1) \qquad (7)$$

Where $i_{sat}$ is the saturation current density of the Schottky contact and $\phi_{m/f}^0$ is the zero-current diode voltage known as the built-in potential in Semiconductor Physics.

In the point defect model we need to invoke, for the proper book-keeping of ion and electron currents, two kinds of channels which feed current into the metal: the defect reactions (1) through (3) or (3') and the Schottky-like pure e-channel. For $\delta = \chi$, however, this purely e-channel is absent, the diode equation does not play any role and the current at the m/f interface is controlled only by the defect reactions (1) through (3) or (3'). For $\delta \neq \chi$, both the channels contribute to the current at the metal/film interface.

**Theoretical Prediction for the Forms of $\phi_{m/f}$ and $\phi_{f/s}$**

Macdonald and co-workers proposed the following forms for $\phi_{m/f}$ and $\phi_{f/s}$ and computed the values of the parameters $\alpha$ and $\beta$ by fitting the EIS data to the model.

$$\phi_{f/s} = \alpha.V + \beta.pH + \phi_{f/s}^0 \tag{8}$$

$$\varepsilon.L = the.potential.drop.across.the.oxide.film.of.thickness.L \tag{9}$$

and $$\phi_{m/f} = (1-\alpha).V - \beta.pH - \phi_{f/s}^0 - \varepsilon.L + \phi_R \tag{10}$$

Let us now discuss the possibility of providing a theoretical basis for these forms which were originally chosen for their simplicity and analytical ease.

Theory demands that we have 2 equations for finding the 2 unknowns $\phi_{m/f}$ and $\phi_{f/s}$.

One equation is provided by

$$\phi_{m/f} + \varepsilon.L + \phi_{f/s} = V_{ap} - \phi_R \tag{11}$$

where $\varepsilon$ is the electrical field inside the oxide film and was taken to be a constant in the original point defect model, whereas, in the variant proposed by the present author[2-4],

$$\varepsilon = \rho_f . i \qquad (12)$$

where $\rho_f$ is the film resistivity and $i$ the current density through the film.

Hence if we use the variant, this equation will be

$$\phi_{m/f} + \rho_f . i . L + \phi_{f/s} = V_{ap} - \phi_R \qquad (13)$$

The second equation to be satisfied is:

$i$ (at the m/f interface) = $i$ (at the f/s interface) $\qquad (14)$

which is essentially equation 5 derived earlier. Now, for the steady state, we obtain, using equations 6 and 7

$$i_{sat} . \left( \exp[F.(\phi_{m/f} - \phi^0_{m/f})/RT] - 1 \right) = (\delta - \chi) . F \left( -J_M^{f/s} + J_i^{f/s} + R_7 \right) \qquad (15)$$

Equation 15 is valid only if $\delta \neq \chi$. When $\delta = \chi$ the purely electronic channel at the m/f interface will be inactive and besides, for $\delta = \chi$, equation 14 will be superfluous as it is already implied by the steady state assumption. Hence, for the case when $\delta = \chi$, we need to find an alternative equation. This equation may be chosen as:

$$2 J_O^{m/f} = \chi . R_7 \qquad (16)$$

which represents the equality of the rate of formation of the new oxide at the m/f interface and its rate of dissolution at the f/s interface.

For the variant proposed by the present author, $J_O^{m/f}$ may be written as $(\chi/2).(J_i^{m/f} - J_M^{m/f})/q$ and equation (16) becomes

$$\left( J_i^{m/f} - J_M^{m/f} \right) = q . R_7 \qquad (17)$$

To summarize, the pair of equations to be solved for finding $\phi_{m/f}$ and $\phi_{f/s}$ are:

In the original PDM

$$\phi_{m/f} + \varepsilon . L + \phi_{f/s} = V_{ap} - \phi_R \qquad (11)$$

and

For $\delta = \chi$, $\quad k_3 = R_7$ (18)

For $\delta \neq \chi$, $\quad i_{sat}.\left(\exp[F.(\phi_{m/f} - \phi^0_{m/f})/RT] - 1\right) = (\delta - \chi).F(k_2 + k_4 + R_7)$ (19)

Where we have used the fact $-J^{f/s}_M = k_4$ and $J^{f/s}_i = k_2$ in equation (15).

In the variant PDM

$$\phi_{m/f} + \rho_f.i.L + \phi_{f/s} = V_{ap} - \phi_R$$ (13)

For $\delta = \chi$, $\quad k_2 + k_4 = q.R_7$ (20)

For $\delta \neq \chi$, $\quad i_{sat}.\left(\exp[F.(\phi_{m/f} - \phi^0_{m/f})/RT] - 1\right) = (\delta - \chi).F.R_{PB}.R_7$ (21)

We have attempted to solve this pair of equations for the above 4 different cases under some assumptions and by employing the Lambert-W function to handle a non-linear equation arising from the pair of equations (13) and (21) and the Results are presented in Table I.

**Discussion and Conclusions**

For the internal consistency of the point defect model, the net current at the m/f interface must equal that at the f/s interface. In this paper this natural constraint is duly imposed on the point defect model and its implications for the functional forms of $\phi_{m/f}$ and $\phi_{f/s}$ are derived. This derivation and the results presented in Table I clearly show that the forms so far assumed by Macdonald and others are valid only under certain conditions and hence one should not blindly use these forms for fitting the experimental data. However it may be remarked that Macdonald et al found the values of the parameters $\alpha$ and $\beta$ to be close to 0.5 and 0.01 respectively by fitting their model to experimental data [5]. These values closely agree with the theoretical predictions of the present work. Table I also presents results which can clearly differentiate between the original PDM and its variant, when applied to relevant steady state and EIS experimental data.

The originators of the point defect model might argue that, though the equality of the currents at the m/f and f/s interfaces is not enforced in the model itself a priori, the fitting of experimental data with theory would find the values of the parameters $\alpha$ and $\beta$ which would ensure this equality.

This is true only in those cases [see Table I] where these parameters $\alpha$ and $\beta$ are linear in $V_{ap}$ and $pH$ , **NOT IN GENERAL.**

The role of the e-channel at the m/f interface is thus clearly established and the kinetics of this e-channel enters the model besides the kinetics of the interfacial defect reactions. In fact, without invoking this e-channel, the currents at the m/f and f/s interfaces will never match for $\delta \neq \chi$ !

The developments reported here are also relevant for the physics of metal-semiconductor contacts as all real Schottky contacts will have defect reactions in addition to pure electron transfer at the interface. Work on the application of the point defect model for metal-semiconductor-metal sandwiches along these lines shall be taken up shortly in our Lab, where the metal takes the place of the electrolyte in the original point defect model and an identical set of defect and electron transfer reactions take place at the two metal-semiconductor interfaces.

The framework presented here is also applicable to non-steady state conditions and in particular for modeling the EIS response of the point defect model. So far Macdonald and co-workers developed the EIS response of the PDM using the assumed forms of $\phi_{m/f}$ and $\phi_{f/s}$. In fact we do not need these forms for deriving the EIS response. One can simply include the pair of equations (13) and (14) in the relevant equations which govern the EIS response and solve the resulting set of equations. As all equations are linear for the small signal ac response **one can thus find the exact impedance response without assuming any forms for $\phi_{m/f}$ and $\phi_{f/s}$.** Indeed the empirical parameters $\alpha$ and $\beta$ of the original PDM would disappear from the exact analytic impedance function. This is a welcome situation as we thus get rid of all empirical parameters from the model. The work on impedance along these lines is in an advanced stage in our Lab and will shortly be reported.

Last, but not the least, the present work provides the correct theoretical basis for apportioning the total applied voltage into the various potential drops in all electrochemical and solid-state systems where a compact layer is sandwiched between a metal and an electrolyte or another metal whose dc or ac responses are sought. Such systems are in abundance in the fields of Electrochemistry and Semiconductor Physics.

## Table I

## Theory-predicted Forms for $\phi_{m/f}$ and $\phi_{f/s}$

### Case A [Original PDM]

(i) $\underline{\delta = \chi}$

$$\phi_{m/f} = -\frac{2.303.n.RT}{\alpha_3.\chi.F} pH + \phi_3^0 + \frac{RT}{\alpha_3.\chi.F}\ln\left(\frac{k_7^0}{k_3^0}\right) - \frac{2.303.n.RT}{\alpha_3.\chi.F}\log C_{H^+}^0 \qquad \text{(T-I)}$$

$$\phi_{f/s} = V_{ap} + \frac{2.303.n.RT}{\alpha_3.\chi.F} pH - \varepsilon.L - \phi_3^0 - \frac{RT}{\alpha_3.\chi.F}\ln\left(\frac{k_7^0}{k_3^0}\right) - \frac{2.303.n.RT}{\alpha_3.\chi.F}\log C_{H^+}^0 + \phi_R \qquad \text{(T-II)}$$

(ii) $\underline{\delta > \chi}$

$$\phi_{m/f} = \frac{V_{ap}}{2} - \frac{2.303.n.RT}{2.F} pH - \frac{\varepsilon.L}{2} - \frac{2.303.RT}{2F}\ln\left(\frac{\left[\overline{i_{sat}} - (\delta-\chi).F.\overline{k_2^0}\right]\left[C_{H^+}^0\right]^n}{(\delta-\chi).F.\overline{k_7^0}}\right) - \frac{\phi_R}{2}$$

... T (III)

and

$$\phi_{f/s} = \frac{V_{ap}}{2} + \frac{2.303.n.RT}{2.F} pH - \frac{\varepsilon.L}{2} + \frac{2.303.RT}{2F}\ln\left(\frac{\left[\overline{i_{sat}} - (\delta-\chi).F.\overline{k_2^0}\right]\left[C_{H^+}^0\right]^n}{(\delta-\chi).F.\overline{k_7^0}}\right) - \frac{\phi_R}{2}$$

... T (IV)

where

$$\overline{i_{sat}} = i_{sat}.\exp[-F\phi_{m/f}^0/RT] \qquad \text{... T (V)}$$

$$\overline{k_2^0} = k_2^0.\exp[-\alpha_2.\chi.F\phi_2^0/RT] \qquad \text{... T (VI)}$$

$$\overline{k_7^0} = k_7^0.\exp[-\alpha_7.(\delta-\chi)..F\phi_7^0/RT] \qquad \text{... T (VII)}$$

Assumptions made: $\alpha_2 = 1/\chi$ ; $\alpha_4 = 1/\delta$ and $\alpha_7 = 1/(\delta-\chi)$.

(iii) $\underline{\delta < \chi}$

In this case too we obtain an exact analytical form for $\phi_{m/f}$ and $\phi_{f/s}$ though they are not linear in $V_{ap}$ and $pH$.

## Case B [Variant PDM]

(i) $\delta = \chi$

In this case an exact analytical form results for $\phi_{m/f}$ and $\phi_{f/s}$ though they are not linear in $V_{ap}$ and $pH$.

(ii) $\delta > \chi$

$$\phi_{m/f} = \frac{\alpha_7(\delta-\chi)}{1+\alpha_7(\delta-\chi)}.V_{ap} - \frac{2.303.n.RT}{F.[1+\alpha_7(\delta-\chi)]}pH + \frac{\alpha_7(\delta-\chi)}{[1+\alpha_7(\delta-\chi)]}\{\alpha_7(\delta-\chi)\phi_7^0 - \phi_R\}$$
$$- \frac{2.303.RT}{F.[1+\alpha_7(\delta-\chi)]}\{n.\log C_{H^+}^0 - \log[(\delta-\chi).F.R_{PB}k_7^0]\} - \frac{2.303.RT}{F}\log \overline{i_{sat}}$$

...T (VIII)

and

$$\phi_{f/s} = \frac{1}{1+\alpha_7(\delta-\chi)}.V_{ap} + \frac{2.303.n.RT}{F.[1+\alpha_7(\delta-\chi)]}pH + \frac{1}{[1+\alpha_7(\delta-\chi)]}\{\alpha_7(\delta-\chi)\phi_7^0 - \phi_R\}$$
$$+ \frac{2.303.RT}{F.[1+\alpha_7(\delta-\chi)]}\{n.\log C_{H^+}^0 - \log[(\delta-\chi).F.R_{PB}k_7^0]\}$$

...Y (IX)

This is an asymptotic result.

(ii) $\delta < \chi$

$$\phi_{m/f} = V_{ap} - \frac{2.303.n.RT}{F.\alpha_7(\delta-\chi)}pH - \phi_7^0 - \frac{2.303.RT}{F.\alpha_7(\delta-\chi)}\{n.\log C_{H^+}^0 - \log[(\chi-\delta).F.R_{PB}k_7^0]\} - \frac{2.303.RT}{F.\alpha_7(\delta-\chi)}\log i_{sat}$$
$$+ \rho_f.L.\frac{\delta}{(\delta-\chi)}.i_{sat} - \phi_R$$

...T (X)

and

$$\phi_{f/s} = \frac{2.303.n.RT}{F.\alpha_7(\delta-\chi)}pH + \phi_7^0 + \frac{2.303.RT}{F.\alpha_7(\delta-\chi)}\{n.\log C_{H^+}^0 - \log[(\chi-\delta).F.R_{PB}k_7^0]\} + \frac{2.303.RT}{F.\alpha_7(\delta-\chi)}\log i_{sat}$$

...T (XI)